# Structurally Dependent Fano Resonances in the Infrared Spectra of Phonons in Few-Layer Graphene

Zhiqiang Li[1*], Chun Hung Lui[1*], Emmanuele Cappelluti[2,3], Lara Benfatto[3,4], Kin Fai Mak[1], G. Larry Carr[5], Jie Shan[6], and Tony F. Heinz[1†]

*1 Departments of Physics and Electrical Engineering, Columbia University, 538 West 120th Street, New York, NY 10027, USA*

*2 Instituto de Ciencia de Materiales de Madrid, CSIC, 28049 Cantoblanco, Madrid, Spain*

*3 Istituto dei Sistemi Complessi, U.O.S. Sapienza, CNR, via dei Taurini 19, 00185 Rome, Italy*

*4 Dipartimento di Fisica, Università "La Sapienza", P.le A. Moro 2, 00185 Rome, Italy*

*5 National Synchrotron Light Source, Brookhaven National Laboratory, Upton, NY 11973, USA*

*6 Department of Physics, Case Western Reserve University, 10900 Euclid Avenue, Cleveland, OH 44106, USA*

*\* These authors contributed equally to this work.*
*† Corresponding author: tony.heinz@columbia.edu*

**Abstract**

The in-plane optical phonons around 200 meV in few-layer graphene are investigated utilizing infrared absorption spectroscopy. The phonon spectra exhibit unusual asymmetric features characteristic of Fano resonances, which depend critically on the layer thickness and stacking order of the sample. The phonon intensities in samples with rhombohedral (ABC) stacking are significantly higher than those with Bernal (AB) stacking. These observations reflect the strong coupling between phonons and interband electronic transitions in these systems and the distinctive variation in the joint density of electronic states in samples of differing thickness and stacking order.

Recent studies have revealed a broad range of novel physical phenomena in few-layer graphene (FLG) [1-8]. The band structure of FLG exhibits a sensitive dependence to both the layer thickness and the stacking sequence of the graphene sheets [1]. It has recently been shown experimentally that FLG samples with Bernal (AB) stacking are semimetals [9], whereas those with rhombohedral (ABC) stacking are semiconductors with tunable band gaps [6]. Furthermore, new phenomena resulting from many-body interactions and Berry's phase are expected in FLG [2-5]. Rhombohedrally (ABC) stacked FLG, for example, has been predicted to exhibit broken symmetry states, including pseudospin magnetism and various spontaneous quantum Hall states [2], associated with the materials' flat bands with diverging density of states (DOS) at low energy and the resulting enhanced electron-electron interactions. In view of these intriguing phenomena, it is of great interest to explore the role of sample thickness and stacking order in the physics of FLG.

In this paper, we report the observation by infrared (IR) spectroscopy of the in-plane optical (G-mode) phonons in FLG samples of differing layer number ($N$ = 3, 4, 5, 6) and crystallographic stacking sequence. In samples of Bernal (AB) stacking and rhombohedral (ABC) stacking, the phonon absorption spectra display the distinctive features of Fano resonances [10] arising from the discrete phonon transition embedded in, and interacting with, the continuum of electronic transitions. Both the line shape and intensity of the spectra reveal a dramatic dependence on the sample thickness and stacking order: The phonon feature displays strongly asymmetric line shapes, as well as ones varying from predominantly increased absorption to increased transmission, depending on the physical structure of the FLG sample. Furthermore, the phonon intensities in ABC-stacked FLG are substantially (~10 times) higher than those in their AB-stacked counterparts. We show that these properties arise from the strong coupling between phonons and electronic excitations in these systems. We provide a theoretical description of our findings in terms of the mixed current-phonon response function recently employed for bilayer graphene [11]. This approach provides a natural means of understanding the relationship between the line shape and intensity of the phonon features, on the one hand, and the electronic structure of the FLG sample on the other. Our work extends previous studies of the Fano resonance in bilayer graphene [12, 13]. In these studies, the material's electronic transitions were tuned with respect to the optical phonons through modification of the band structure and doping level [12, 13]. For FLG, such diverse resonant behavior is achieved naturally through the intrinsic electronic properties of FLG of differing physical structure.

We investigated few-layer graphene samples deposited on $SiO_2$/Si substrates by mechanical exfoliation of kish graphite. IR reflectance and transmission measurements were performed using a Fourier Transform IR spectrometer coupled with an IR microscope in the range of photon energy (*E*) of 0.1 -1 eV with a spectral resolution of 0.25 meV. The number of layers for a given sample could be precisely determined from transmission measurements in the near-IR range (> 0.8 eV), where each graphene monolayer absorbs approximately 2.3% of the light [14-17]. From the reflectance and transmission data, the complex optical conductivity spectra, $\sigma_1(E)+i\sigma_2(E)$, of few-layer graphene were extracted. The analysis takes into account the multiple reflections in the substrate [18].

In samples with three to six layers, two distinct types of absorption spectra $\sigma_1(E)$ were observed for samples with exactly the same number of layers. One class of spectra follows the results previously reported for Bernal stacking [15]. The other class is characterized by two significantly stronger absorption peaks, with positions that shift successively to lower energy with increasing layer number (Fig. 1 and Fig.S1) [19]. Previous studies [8, 20] have shown that samples with rhombohedral stacking exhibit such pronounced absorption peaks, which arise because of the high electronic joint density of states near the critical points in the band structure. The evolution of the band structure of rhombohedral samples is also predicted to produce a red shift in the low-energy absorption peaks with increasing layer number [21]. The agreement between the second class of spectra and the above expectations indicates the dominance of rhombohedral stacking order in these samples [22].

In addition to the broad features in the absorption spectra from electronic transitions, we also observed (Fig. 1) pronounced narrow features at ~ 200 meV, the energy of G-mode (in-plane, Γ-point optical) phonons [23]. However, instead of a symmetric resonance feature corresponding to increased absorption that would normally be expected for excitation of these modes, we find both resonance and anti-resonance (corresponding to reduced absorption) line shapes, as well as dispersion line shapes. The line shape and intensity of the phonon spectra exhibit striking changes as a function of layer thickness and stacking order. In particular, the phonon features in the rhombohedral samples show much (~ 10×) higher intensity than those in the Bernal samples.

The data in Fig. 1 were measured from similarly prepared samples with comparable carrier densities from unintentional doping, which are estimated to be around $5 \times 10^{12}$ cm$^{-2}$ [19]. To check the effect of doping on the phonon features, we performed optical measurements on trilayer samples with electrical gating [6]. We found that while the phonon spectra of samples with either stacking order can be continuously tuned by gating, the phonon intensity of ABC samples is always much higher than that of ABA samples at similar doping levels (except at low doping levels where charge inhomogeneity can complicate the intrinsic physics).

The observed resonance, anti-resonance and dispersion absorption line shapes are characteristic of Fano resonances arising from coupling between the phonons and an electronic continuum [10]. Therefore, in the vicinity of the phonon energy we described the experimental $\sigma_t(E)$ spectra as $\sigma_t(E) = \sigma_\varepsilon(E) + \sigma_{Fano}(E)$. Here we subtract away the background electronic absorption $\sigma_\varepsilon(E)$ [19] to highlight the phonon feature. We then compare the result to the Fano resonance line shape described by [10]:

$$\sigma_{Fano}(E) = \frac{2p}{\pi\Gamma(q^2+1)} \frac{(q+\varepsilon)^2}{(1+\varepsilon^2)}, \qquad (1)$$

where $\varepsilon = 2(E - E_{ph})/\Gamma$, $E_{ph}$ and $\Gamma$ are, respectively, the energy and width of the phonon feature. In this expression, $p$ denotes the phonon intensity parameter, which is always positive and vanishes when the phonon feature is absent [11]. The Fano parameter $q$, which is a real number [24], governs the line shape. When $|q| \gg 1$, we can consider the phonon to dominate the response and a resonance is observed (Eq. 1 approaches a Lorentzian); on the other hand, when $|q| \ll 1$, electronic excitation dominates and an anti-resonance feature is observed. For intermediate values of $|q| \sim 1$, the electronic and phonon contributions are comparable and a dispersion line shape is present. In our treatment, we consider $p$ and $q$ to be constants over the narrow energy region of the phonon. The experimental data can be well described by Eq. (1) and a broad electronic background. Fitting parameters are shown in Fig. 2 and Fig. S3 [19].

We note that in *N*-layer graphene the weak interlayer coupling leads to the presence of *N* distinct modes for the G-mode phonon. However, these splittings are quite small, < 1 meV [25], and cannot be resolved spectrally due to the finite width of the phonons. In our fit, we accordingly treat the observed feature as a single phonon. The parameters from the fit are thus phenomenological quantities representing the combined effect of all IR active phonons. We note that in contrast to the IR spectra presented here, the Raman spectra of the G-mode phonon in all the few-layer samples do not exhibit significant Fano resonance characteristics. We attribute this result to the weakness of possible electronic Raman processes, which are necessary to produce the continuum signal for a Fano resonance.

We now consider the origin of the strikingly different spectra, and, correspondingly, different phonon strength *p* and Fano parameter *q*, for the resonances in FLG of different thickness and stacking order. The observed behavior can be explained qualitatively by the Fano theory, in which parameters *p* and *q* can be related to the optical transition amplitudes of the phonon and electronic continuum, $A_{ph}$ and $A_e$, and the electron-phonon coupling strength $V_{e\text{-}ph}$: $q = A_{ph}/\pi V_{e\text{-}ph} A_e$ [10] and $p = |A_{ph}|^2 (1 + 1/q^2)$ [10, 11]. In nonpolar materials such as graphene, the optical transition amplitude of a phonon is determined largely by its coupling to the electronic transitions. This coupling, mediated by electron-phonon scattering, reflects the electronic density of states. Thus, to obtain a qualitative understanding of the observed trends, we use the approximation $V_{e\text{-}ph}(E) A_e(E) \sim |A_e(E)|^2$. One can then deduce [19] an expression for the resonant optical transition amplitude of phonon of

$$A_{ph}(E_{ph}) = P\int \frac{dE}{E_{ph} - E} |A_e(E)|^2 . \tag{2a}$$

Here P denotes the principal value of the integral. Within the same approximation, the Fano line shape parameter *q* is given by

$$q = \frac{1}{\pi} P\int \frac{dE}{E_{ph} - E} \left| \frac{A_e(E)}{A_e(E_{ph})} \right|^2 . \tag{2b}$$

Eqs. (2a, b) relate the characteristics of the Fano resonance to disposition of the distribution of transition strengths of the electronic continuum $|A_e(E)|^2$ with respect to the phonon energy $E_{ph}$. Since the weighting of the continuum transitions drops off as the energy moves away from that of the phonon, we expect that the dominant behavior will be determined by the influence of the low-energy interband transition peak (Fig. 1).

This picture offers a natural explanation for the evolution of the Fano parameter $q$ with layer thickness. For rhombohedral FLG samples with three and four layers, the phonon energy lies rather far below the center of the electronic resonance, and $|A_e(E)/A_e(E_{ph})| \gg 1$ for most parts of the spectrum. Then Eq. (2b) predicts $|q| \gg 1$. For five-layer rhombohedral samples, the phonon lies in the shoulder of the strong electronic transition peak. The large value of $|A_e(E_{ph})|^2$ therefore leads to a significant reduction of the value of $q$ and, correspondingly, to the observation of a strongly asymmetric phonon feature. Interestingly, the phonon lies near the center of the low-energy electronic resonance in rhombohedral six-layer graphene. In this case, cancellation of the contributions from low ($E < E_{ph}$) and high energy ($E > E_{ph}$) sides of the phonon in the integral in Eq. (2b) leads to a rather small value of $q$ (~ 0.2) and a dip feature in the phonon spectrum. Similarly, the phonons in Bernal samples always lie at the low energy shoulder of the broad electronic resonance peaks (Fig. 1). This yields $|q| \sim 1$ and dispersion-shaped features in the phonon spectra, analogous to the rhombohedral five-layer samples.

The substantial dependence of the phonon intensity $p = |A_{ph}|^2 (1 + 1/q^2)$ on the stacking order and layer thickness can also be explained qualitatively within this picture. Based on Eqs. (2a, b), the higher phonon intensity in rhombohedral samples arises from their stronger electronic absorption peaks compared to those in Bernal samples. Moreover, the increase of $p/N$ with layer thickness $N$ in rhombohedral samples can be understood as follows: The electronic resonance shifts downwards in energy with increasing $N$ and approaches the phonon energy $E_{ph}$, leading to larger value of $1/(E_{ph} - E)$ in Eqs. (2a, b) and thus greater values of $p/N$. The slight drop of $p/N$ for rhombohedral six-layer samples stems from the reduction of $A_{ph}(E_{ph})$ because of the cancellation effects discussed above. Therefore, the trends for both the phonon line shape (through $q$) and phonon intensity (through $p$) can be understood based on the coupling between phonons and low-energy interband transitions.

In order to explain our observations within a more quantitative framework, we compute within the tight-binding model the mixed current-phonon response function $\chi_{j\nu}(E)$ [11, 19]. This response function can be expressed for each phonon mode $\nu$ as $Re\ \chi_{j\nu}(E_{ph}) = A_{ph}$ and $Im\ \chi_{j\nu}(E_{ph}) = -\pi V_{e\text{-}ph} A_e$. One can then compute the intensity of the infrared active mode $\nu$ as $p_\nu = |\chi_{j\nu}(E_{ph})|^2$ and the Fano parameter as $q_\nu = -Re\ \chi_{j\nu}(E_{ph})\ /\ Im\ \chi_{j\nu}(E_{ph})$. We find in the rhombohedral 6-layer samples and all Bernal samples that several phonon modes are involved in the phonon spectra with comparable intensity [19]. This prevents a direct comparison between our theoretical calculations and the Fano analysis in Fig. 1, which assumes the presence of only one effective mode. On the other hand, the rhombohedral samples with $N \leq 5$ show a single dominant IR active phonon mode $\bar{\nu}$, which is an antisymmetric mode with the carbon atoms in the $i$-th ($i =1, 2, \ldots N$) layer moving exactly out of phase with respect to those of the same sublattice in the ($N+1-i$)-th layer. Therefore, the Fano parameter $q$ and phonon intensity $p$ in these samples are well-defined within our calculations. As shown in Fig. 2, the evolution of $q$ and $p$ with the number of layers is qualitatively reproduced by our calculations. The disagreement between experimental and theoretical values may reflect the effect of additional interlayer coupling parameters $\gamma_i$ ($i = 2,3,4,5$), which have been omitted in our model for simplicity.

To explain the substantial difference between rhombohedral and Bernal samples, in Fig. 3 we show the electronic band structure and density of states (DOS) for 5-layer graphene with the two different stacking orderings. In rhombohedral samples the high DOS points in different bands occur roughly at the same momentum **k**, leading to singularities in the joint-DOS and very pronounced peaks in the optical absorption spectra [8]. Moreover, the resonance energy for the lowest interband transition is close to the phonon frequency $E_{ph}$. These features give rise to substantial spectral weight in the mixed response functions and therefore strong IR phonon intensities in rhombohedral samples (Fig. 3). In particular, the above mentioned antisymmetric mode in rhombohedral samples ($N < 6$) shows the strongest coupling to the low-energy interband transition, therefore this mode acquires dominant IR intensity. On the other hand, the absence of singularities in the joint-DOS in Bernal samples [8] results in a weaker IR intensity of phonons. An additional factor contributing to the relative weakness of the phonon feature in Bernal FLG is the presence of several IR active phonons in these materials [19]. Since the different phonons have somewhat different frequencies and may have different values of the corresponding Fano parameter $q$, the observed spectrum arising from the sum of several such asymmetric lines is expected to exhibit cancellation effects.

In summary, we have observed features associated with the zone-center optical phonons in FLG of both Bernal and rhombohedral stacking. The phonon features exhibit Fano resonance line shapes that arise from coupling to the electronic continuum. Our study has revealed two important degrees of freedom in defining the characteristics of the Fano response: The layer thickness and crystallographic stacking sequence. This leads to a wide diversity in the observed line shape for different FLG samples without any external modification of properties. The significantly enhanced phonon intensity in rhombohedrally stacked FLG compared to that in the conventional Bernal stacked FLG reflects the distinctive electronic structure of these materials and the corresponding presence of strong, sharply defined low-energy electronic transitions.

This work was supported by the National Science Foundation under grant DMR-0907477 at Case Western Reserve and under grants DMR-1106225 and CHE-0641523 at Columbia, with additional funding from NYSTAR. E.C. acknowledges support from the European FP7 Marie Curie project PIEF-GA-2009-251904. The IR measurements were performed at the National Synchrotron Light Source, which is funded by the U.S. DOE under contract DE-AC02-98CH10886.

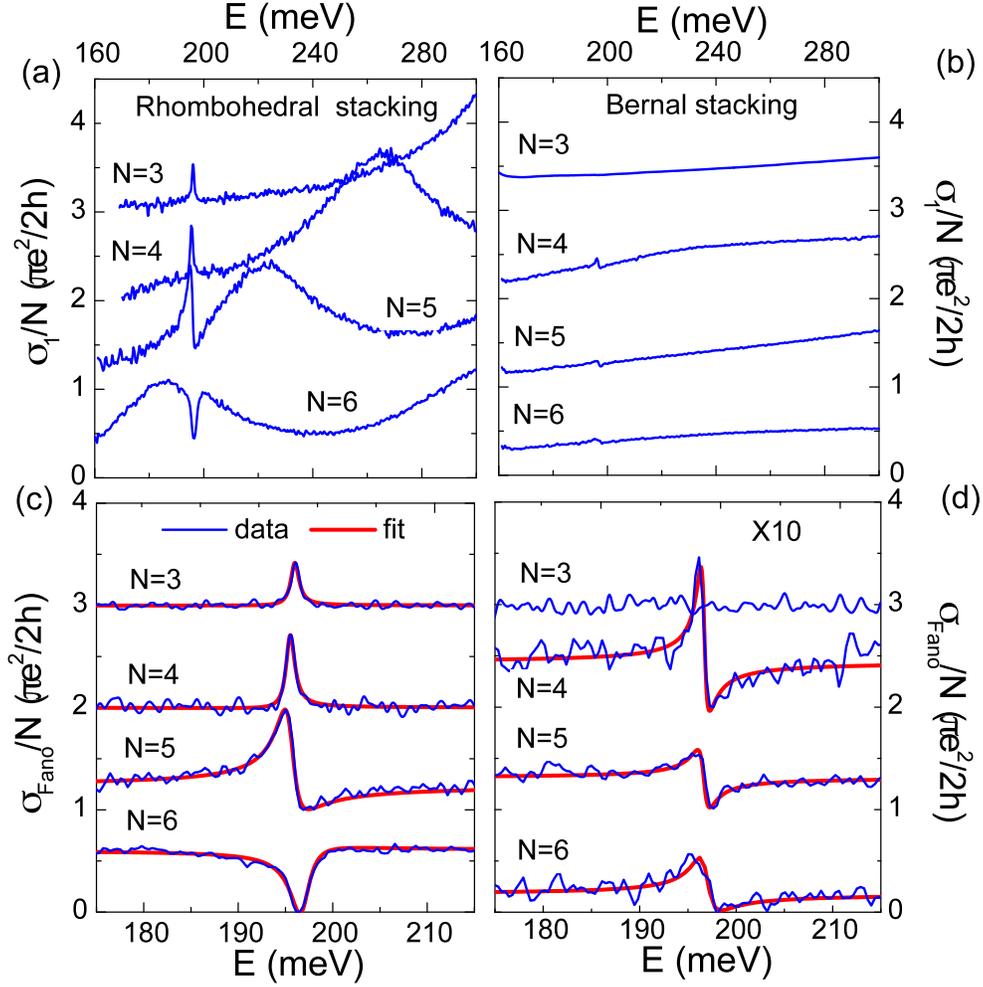

Figure 1: (a-b) The normalized optical conductivity, $\sigma_1(E)/N$, of $N$-layer graphene with rhombohedral and Bernal stacking. (c-d) The normalized Fano spectra $\sigma_{Fano}(E)/N$ in the region of the phonon feature for FLG with rhombohedral (c) and Bernal (d) stacking, obtained by subtraction of an electronic background from the data in (a) and (b). The red lines are fits using Eq. (1). The spectra in (d) have been multiplied by a factor of 10. For clarity, the spectra are displaced vertically from one another by increments of $(\pi e^2/2h)$.

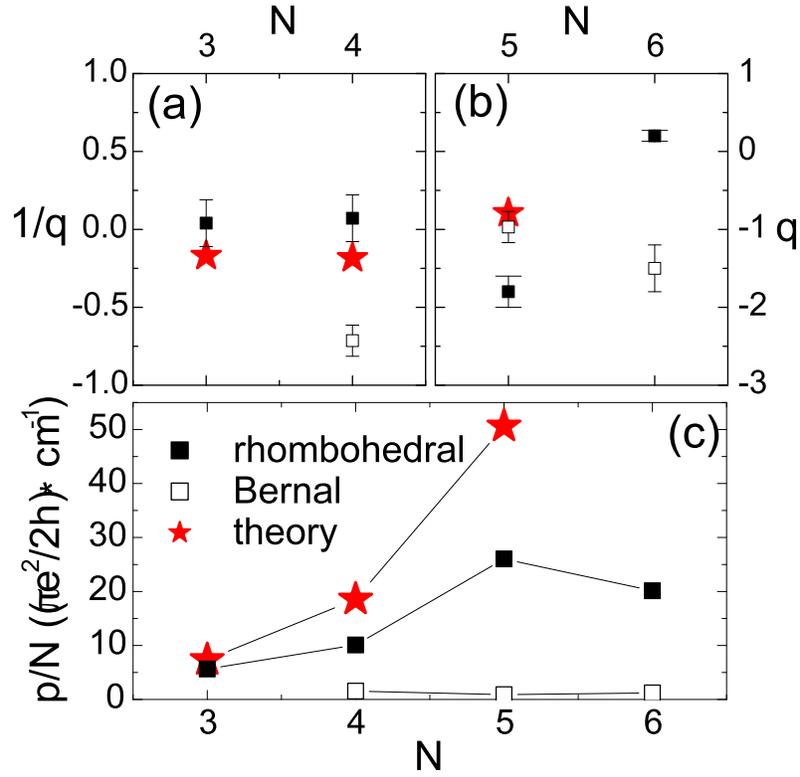

Figure 2: (a-b) The Fano parameters $1/q$ ($q$) for Bernal and rhombohedral samples obtained from the fits. The error bars reflect the experimental accuracy and the uncertainty from subtraction of the background. (c) The phonon intensity normalized by layer number $p/N$. Theoretical values for rhombohedral samples evaluated using the mixed response function are also shown.

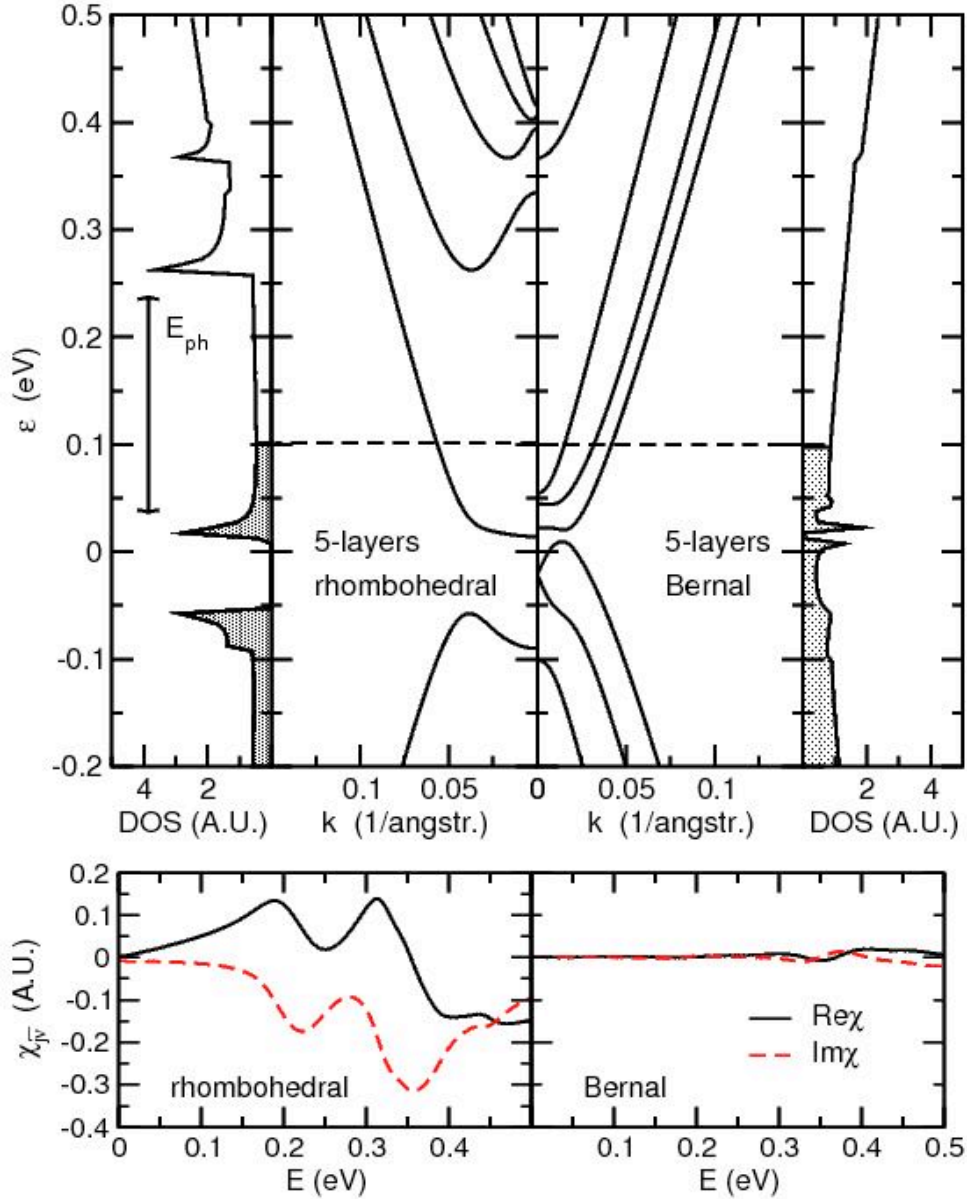

Figure 3: Upper panels: A representation of the electronic band structure and DOS for 5-layer graphene samples with rhombohedral and Bernal stacking. The dashed line represents the chemical potential, shifted from charge neutrality because of unintentional doping. Lower panels: The real and imaginary part of the corresponding mixed current-phonon response function responsible for the IR activity.

# Supplementary Material

**1. Data for samples with rhombohedral stacking**

Figure S1 displays the same data as those in Fig. 1(a) of the main text for the optical conductivity spectra of the rhombohedrally stacked samples, but over a broader range of energy. The conductivity spectra show two pronounced broad resonances that shift to low energy as the layer thickness is increased, as expected for ABC-stacked FLG [1].

The band gap in the ABC trilayer sample ($N = 3$) in Fig. S1 is around 75 meV, as indicated by the separation of the two resonances in the conductivity spectrum [2]. Based on electrical gating experiments, this gap value corresponds to a carrier concentration of around 5 x$10^{12}$ cm$^{-2}$ [2]. All of the samples in Fig. 1 of the main text were prepared using similar procedures and conditions by simple exfoliation. They are expected to exhibit comparable carrier densities from unintentional doping, which is understood to be determined primarily by the SiO$_2$ substrate.

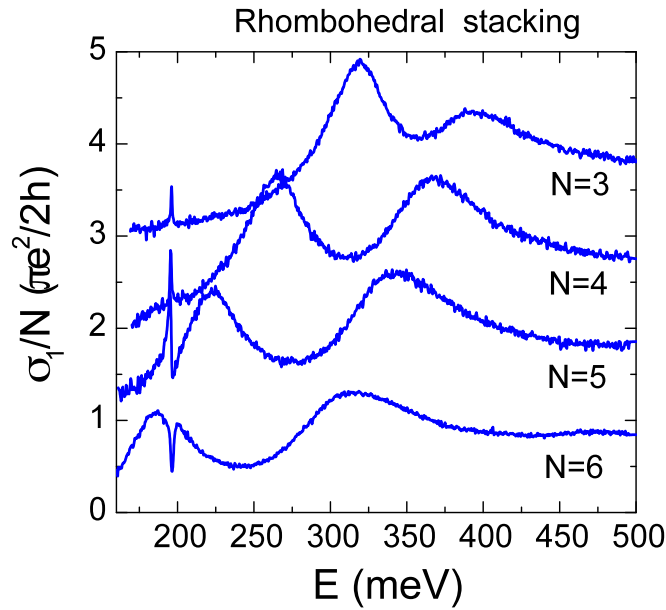

Figure S1: The normalized optical conductivity spectra, $\sigma_1(E)/N$, for $N$-layer graphene with rhombohedral stacking. Successive spectra are shifted vertically by ($\pi e^2/2h$) from one another.

## 2. Details on data analysis using the Fano theory

In this section, we provide a more detailed discussion of our data within the context of Fano theory [3]. To analyze the observed phonon spectra, we phenomenologically fit the measured $\sigma_1(E)$ spectra in the vicinity of the phonon with one Fano oscillator $\sigma_{Fano}$ (defined in the main text) and two broad Lorentzian oscillators. The latter two oscillators account for the broad electronic background. This fitting procedure is unambiguous if one only aims to discuss the narrow phonon spectrum $\sigma_{Fano}$. A representative fit is shown in Fig. S2 for 6-layer sample with rhombohedral stacking. The parameters obtained from this procedure for the different FLG samples are shown in Fig. 2 of the main text and Fig. S3 below.

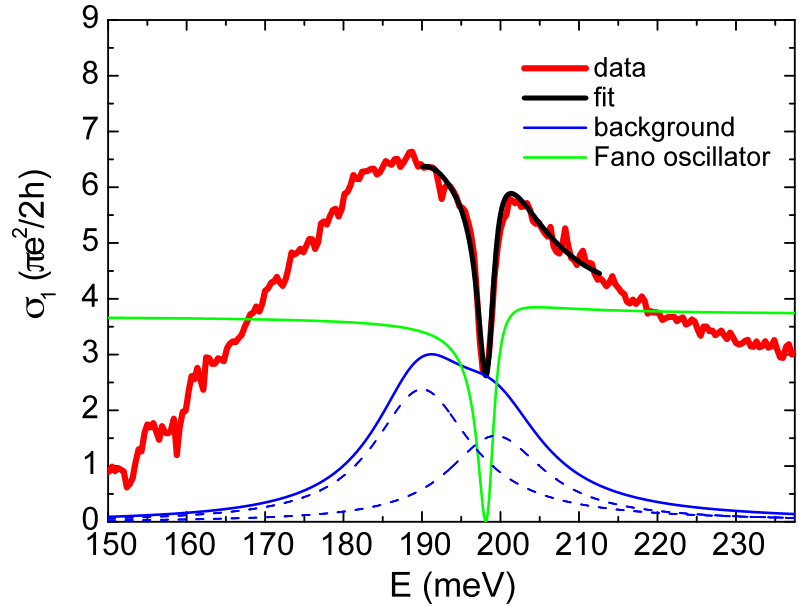

Figure S2: The IR conductivity spectrum $\sigma_1(E)$ for a 6-layer sample with rhombohedral stacking (red) and a representative fit (black) in the vicinity of the phonon using one Fano oscillator $\sigma_{Fano}$ (green curve) and an electronic background (solid blue line) consisting of two Lorentzian oscillators (dashed blue lines). The Fano and Lorentzian oscillators outside the energy range of the fit were evaluated using the parameters obtained from the fit.

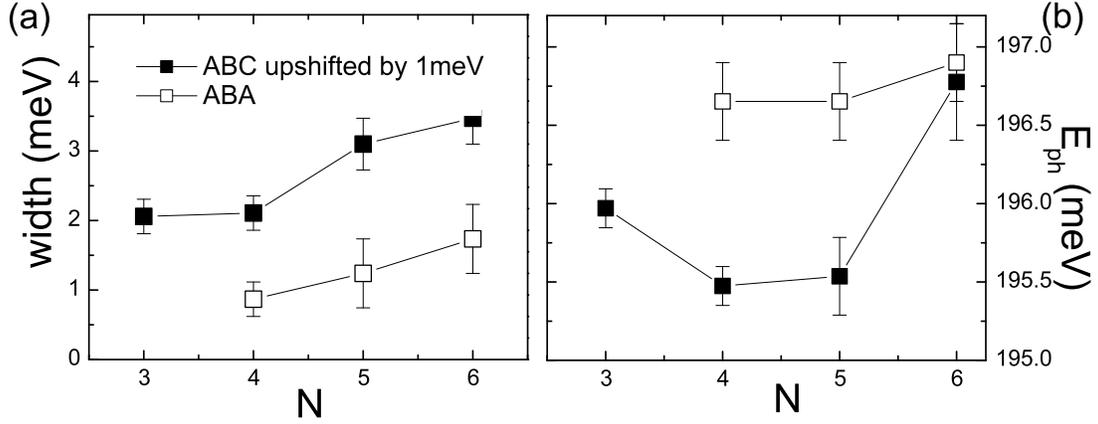

Figure S3: The phonon width (a) and phonon energy (b) obtained from fitting the data in Fig. 1 of the main text using the Fano formula.

We now discuss some relevant issues related to the Fano analysis. In Eq. (1) of the main text, the Fano parameter is defined as [3]:

$$q = \frac{\langle \Phi|T|i\rangle}{\pi V_{e-ph}\langle \psi_{E_{ph}}|T|i\rangle} = \frac{\langle \varphi|T|i\rangle + P\int dE V_{e-ph}(E)\langle \psi_E|T|i\rangle/(E_{ph}-E)}{\pi V_{e-ph}\langle \psi_{E_{ph}}|T|i\rangle}. \quad (S1)$$

Here $V_{e-ph}$ is the coupling strength between the bare phonon $\varphi$ and the electronic continuum $\psi_E$, and $\Phi$ is the modified phonon. P represents the principal part of the integral. $T$ is a transition operator and $\langle X|T|i\rangle$ denotes the transition matrix element from an initial state $|i\rangle$ to a final state $|X\rangle$. We define: $A_e = \langle \psi_E|T|i\rangle$ and $A_{ph} = \langle \Phi|T|i\rangle$. We then have $q = A_{ph}/\pi V_{e-ph} A_e$ and $p = |A_{ph}|^2 (1+1/q^2)$.

In this context, the trends of the Fano parameter $q$ in our data can be understood from a close examination of Eq. (S1). The intensity of the modified phonon $\langle \Phi|T|i\rangle$ is characterized by the total effective polarization charge of the carbon atoms, which is on the order of 1 per atom in all samples estimated from the magnitude of the Fano resonance, as first discussed in [4]. On the other hand, the intensity of the bare phonon $\langle \varphi|T|i\rangle$ is determined by the polarization charge induced by the charge imbalance in the two sublattices, which is less than the free carrier density in the sample ($10^{-3}$-$10^{-2}$ per atom) and thus negligible compared to $\langle \Phi |T|i \rangle$. Therefore, $A_{ph}$ is mainly determined by the coupling of the phonon to the electronic transitions $A_e$ [3]: $A_{ph}(E_{ph}) = P\int dE\, V_{e-ph}(E) A_e(E)/(E_{ph} - E)$. The electron-phonon coupling function $V_{e-ph}(E) A_e(E)$ and the electronic response $|A_e(E)|^2$ exhibit qualitatively similar spectral features that are governed by the electronic density of states [5]. Therefore, in order to acquire a qualitative understanding, we apply the approximation $V_{e-ph}(E) = A_e^*(E)$ to obtain $q = A_{ph}/\pi |A_e|^2$ and $A_{ph}(E_{ph}) = P\int dE\, |A_e(E)|^2/(E_{ph} - E)$. We observe that the peak of the low-energy interband transition lies in the vicinity of the phonon, whereas other resonances are much further away from the phonon in energy. Therefore, the nearby low-energy interband transition, corresponding to large values of the $1/(E_{ph} - E)$ factor, is expected to be the dominant contribution to $q$. Based on the above considerations, we show in the main text that the trends for the intensity and line shape of the Fano resonances can be understood in terms of the characteristics of the low-energy electronic structure of the different FLG samples.

## 3. Theoretical calculation of the electronic structure and mixed response function

In our calculations, we generalize the phenomenological model for the lattice dynamics introduced in [6] for trilayer graphene to the case of $N$-layer samples. Within this model we can obtain the phonon frequencies and phonon eigenvectors from the phonon Hamiltonian:

$$\begin{pmatrix} E_0 & \varepsilon & 0 & \ldots & 0 & 0 \\ \varepsilon & \delta + E_0 & \varepsilon & \ldots & 0 & 0 \\ 0 & \varepsilon & \delta + E_0 & \ldots & 0 & 0 \\ \ldots & \ldots & \ldots & \ldots & \ldots & \ldots \\ 0 & 0 & 0 & \ldots & \delta + E_0 & \varepsilon \\ 0 & 0 & 0 & \ldots & \varepsilon & E_0 \end{pmatrix}$$

where the values of the phonon energy of monolayer graphene $E_0 = 1589.5$ cm$^{-1}$, the phonon interaction strength between adjacent layers $\varepsilon \approx -2$ cm$^{-1}$ and the on-site energy in the middle layers $\delta = 3$ cm$^{-1}$ have been estimated, for both rhombohedral and Bernal stacking, from a comparison with density functional theory (DFT) calculations [6]. They are assumed to be the same for $N$-layer graphene ($N > 3$). A classification of the eigenmodes according to the group theory can be found in [7]. Although the eigenvectors obtained in this way are not expected to be the true eigenstates in the presence of the potential gradient induced by the substrate, they can provide a useful basis for investigation of the relevance of each mode.

In order to determine the electronic structure and the infrared properties, we employ a tight-binding model for $N$-layer graphene, with different stacking orderings taking into account. For simplicity, we take into account only the hopping terms $\gamma_0'=3.12$ eV and $\gamma_1'=0.377$ eV terms [8]. We also consider the influence of the unintentional doping of the samples. The electronic Hamiltonian is thus completed by the knowledge of the total amount of charge doping density $n_0$ in the FLG and an external electric field $E_z$ perpendicular to the layers. In our as-prepared (ungated) samples, $E_z$ arises from charged impurities in the substrate and other unintentional doping, via $E_z=n_{imp}/e_g$, where $n_{imp}$ is the density of the charged impurities and $e_g$ is the out-of-pane dielectric constant of the graphene layers. The exact relation between $n_{imp}$ and $n_0$ presumably depends on the details of the screening [9]. For our present purposes, we consider $n_0$ and $E_z$ as two independent quantities; we estimate them from fitting the experimental optical conductivity spectra (Fig. S1) using theoretical ones calculated with Kubo formula. In a similar fashion as in Ref. [10], the chemical potential $\mu$ and charge distribution in each layer, as well as the potential gradient between different layers, are determined in a self-consistent way starting from a null gradient and iterating until convergence is achieved. For rhombohedral samples, we find: $n_0 = 5\times10^{12}$ cm$^{-2}$ and $E_z = 18$ meV/Å for $N=3$; $n_0 = 7\times10^{12}$ cm$^{-2}$ and $E_z = 27$ meV/Å for $N = 4$; $n_0 =10\times10^{12}$ cm$^{-2}$ and $E_z = 27$ meV/Å for $N=5$. The self-consistent solutions for 5-layer system for both rhombohedral and Bernal stacking order are shown in Fig. S4. The presence of such potential gradient is reflected in the gap-like modification of the electronic dispersion (see Fig. 3 in the main text) and in a finite IR activity for any phonon mode ν. For the rhombohedral stacking with $N < 6$, however, the $\bar{\nu}$ mode is dominant, which is an antisymmetric mode with the carbon atoms in the $i$-th ($i =1, 2, …N$) layer moving exactly out of phase with respect to those of the same sublattice in the ($N+1−i$)-th layer.

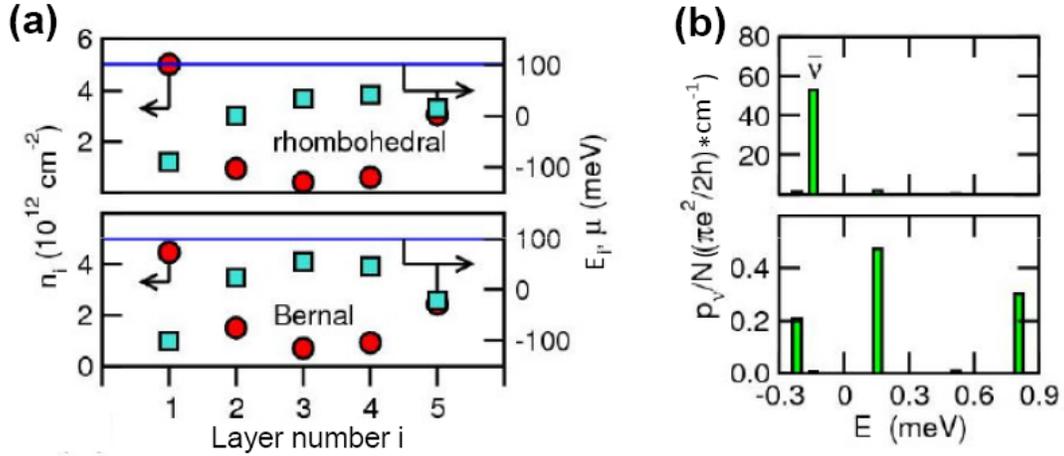

Figure S4: (a) Doping density $n_i$ for each layer (red circles, left side scale), local electrostatic potential $E_i$ (light blue squares, right side scale) and chemical potential $\mu$ (blue solid line, right side scale) for 5-layer rhombohedral and Bernal graphene. (b) Corresponding phonon intensity distribution $p_\nu$ for each mode $\nu$. Note the different vertical scale for the Bernal and rhombohedral stacking. The mode $\bar{\nu}$ can be seen to be dominant for rhombohedrally stacked graphene.

Once the electronic properties of the Hamiltonian have been determined self-consistently by the above procedure, the mixed response function $\chi_{j\nu}(E)$ is computed along the lines described in Ref. [5]. Note that neither the current operator $j$ nor the electron-phonon Hamiltonian $H_\nu$ is affected by the presence of the potential gradient.

## Supplementary References